\documentclass[twocolumn,showpacs,floatfix,prl]{revtex4}

\usepackage[dvips]{epsfig}

\usepackage{float}

\begin{document}

\title{Electron Interactions and Scaling Relations for \\
Optical Excitations in Carbon Nanotubes}

\author{C.L. Kane, E.J. Mele}
\affiliation{Dept. of Physics and Astronomy, University of Pennsylvania, Philadelphia, PA 19104}

\begin{abstract}
Recent fluorescence spectroscopy experiments on
single wall carbon nanotubes reveal substantial
deviations  of observed absorption and emission energies from
predictions of noninteracting models of  the electronic structure.
Nonetheless, the data for nearly armchair nanotubes
 obey a nonlinear scaling relation as a function the tube radius $R$.  We show that these effects can be understood in a
theory of large radius tubes, derived from the theory of two
dimensional graphene where the coulomb interaction leads to a
logarithmic correction to the electronic self energy and marginal
Fermi liquid behavior. Interactions on length scales larger than
the tube circumference lead to strong self energy and excitonic
effects that compete and nearly cancel so that the observed
optical transitions are dominated by the graphene self energy
effects.

\end{abstract}

\pacs{78.67.C, 71.35, 31.25.J}
\maketitle

 The optical transition energies of semiconducting nanotubes, along
with their dependence on the nanotube diameter and chiral angle
have been studied in a recent series of fluorescence spectroscopy
experiments \cite{bachillo,lefebvre,hagen}. Though the
experiments were originally interpreted in the context of a
simple non interacting electron model, it has become increasingly clear that electron
interactions play an important role in determining the optical
transition energies\cite{ando,kanemele,spaturu,tersoff}. As pointed out
in early work by Ando\cite{ando}, interactions lead to (1) an
increase in the single particle energy gap and (2) binding of
electrons and holes into excitons.  More recently, Spataru et
al.\cite{spaturu} have reached a similar conclusion by computing
the optical spectra for selected small radius nanotubes. However,
the systematic dependence of the transition energies on nanotube
radius has not been  addressed.

In this paper we examine the optical excitations of carbon
nanotubes in the limit of large radius, $R$,  where they inherit
their electronic structure from that of an ideal sheet of two
dimensional (2D) graphene.  This permits a systematic study of the
radius and subband dependence of the excitations to leading order
in $1/R$ . In this limit the electron interactions fall into two
categories: (1) 1D interactions on scales longer
than the tube circumference, and (2) 2D interactions
on scales smaller than the tube circumference. We find that the 1D
 long range interaction (1) leads to both a
substantial enhancement of the energy gap and a large exciton
binding energy which both scale as $1/R$. Although both effects
are large they have opposite sign and ultimately lead to a
moderate enhancement of the predicted optical transition energy.
By contrast, we find the 2D interactions (2) lead to
a $\log R/R$ correction to the bandgap renormalization.  This
singular behavior can be traced to the effect of a the Coulomb
interaction on the dispersion of 2D graphene, which
leads to marginal Fermi liquid behavior\cite{guinea}.  This
logarithmic correction is not cancelled by the exciton binding
energy, and leads to a nonlinear scaling dependence of the
transition energies on $R$. The presently available optical data
indeed show this non linear scaling behavior and agrees favorably
with the predictions of the large radius theory even for tubes
with moderately small radii $R \sim 0.5$ nm.

Below we review the non interacting electron predictions for the
energy gaps of semiconducting tubes and show that
they can not
explain the nonlinear scaling behavior present in the observed
transition energies.
We then present the theory for
large radius tubes, focusing first on the effect of the 2D
 interaction on scales shorter than the
circumference.  We then incorporate the longer range 1D
 interactions into the theory.

The simplest model of nanotube electronic structure, based
on non interacting electrons in a linear graphene spectrum, predicts
that the energy gaps of semiconducting nanotubes are
\begin{equation}
E_n^0(R) = 2 n \hbar v_F/3 R,
\end{equation}
where $R$ is the nanotube radius, $n =1, 2$, $4$, $5$ describes
the 1st, 2nd, 3rd and 4th subbands, and  $v_F$ is the graphene Fermi
velocity.  For a tight binding model on a honeycomb lattice with
with lattice constant $a$ and a nearest neighbor hopping amplitude
$\gamma_0$,  $\hbar v_F = \sqrt{3}\gamma_0 a/2$.
% Typically, $v_F$ (or equivalently $\gamma_0$) are fit to experiment.
The linearized
model (1) is exact in the limit of large radius, and is the first
term in an expansion in powers of $1/R$. Corrections due to
curvature\cite{kmcurvature} and trigonal warping\cite{reich} are
proportional to $\nu\sin 3\theta/R^2$, where $\theta$ is the
chiral angle  ($\theta=0$ denotes an armchair wrapping) and
$\nu = \pm 1$ is the chiral index. A central prediction of the non
interacting model is thus that for large $R$ the
band gaps scale linearly with $n/R$ -  a fact that can be traced
to the linear dispersion of graphene at low energies.  The large
$R$ limit is most accurate for nearly armchair nanotubes
for which the $\sin 3\theta$ corrections are smallest.  For such
tubes Eq. 1, describes the tight binding energy gaps to better
than 1\% for tubes with radii as small as $0.5$ nm.  The next term
in the expansion at ${\cal O} (1/R^3)$ is negligible.  Here
we focus exclusively on nearly armchair nanotubes, where large
$R$ scaling can be meaningfully applied.  $\sin 3\theta$
corrections, when present in specific nanotubes, lead to
deviations from the scaling predictions\cite{spaturu,tersoff}.

The observed transition energies do not obey this linear scaling
behavior.  Extrapolated to large radius, the ratio between the
first two transition energies appears to saturate at a value of
1.7, rather than 2 - a fact we have called the ``ratio problem"
\cite{kanemele}.  In addition, the observed transition energies
are systematically larger than the non interacting prediction.  A
nearly armchair nanotube of radius $0.5$ nm has an observed first
subband gap of $0.98$ eV, which corresponds to Fermi velocity of
7.35 eV\AA. This value is significantly larger than the value 5.3
eV\AA\ found in graphite and the value 6.1 eV\AA\
deduced from resonance Raman data.  The ``blue shift" cannot be
represented by a simple scaling of the transition energies since
it is larger for larger radius tubes and therefore not linearly
proportional to $n/R$.

\begin{figure}
 \centerline{ \epsfig{figure=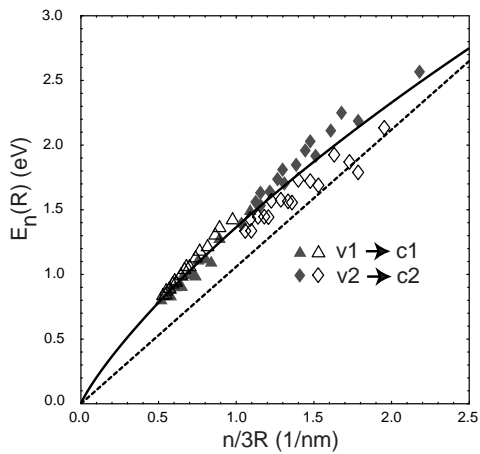,width=2.5in} }
 \caption{Optical transition energies in the first two subbands
 for semiconducting nanotubes measured in Ref. 1
 as a function of $n/3R$.  The filled/open symbols correspond to $[p,q]$ nanotubes
 with chiral index $\nu = p-q \ {\rm mod}\ 3 = +/- 1$.  The dashed line is prediction
 of the non interacting theory.  The solid line is Eq. (4), which
 incorporates the effect of the 2D Coulomb interaction.}
 \end{figure}

In Fig. 1 we plot the transition energies reported in Ref. 1 as a
function of $n/3 R$, where $n$ is the subband index, and $R$ is
the tube radius deduced in Ref. 1 by exploiting the pattern of
$\sin 3\theta/R^2$ corrections.  We have used different symbols to
represent the data with $\nu=\pm 1$. The separatrix
between the data for the positive and negative $\nu$ locates the
data for the nearly armchair tubes with $\theta \sim 0$. At the
separatrix the $\sin 3\theta/R^2$ corrections are absent, so that
the radius dependence should be described by the large $R$ limit
to order ${\cal O} (1/R^3)$ .  It is clear, however, that even at
the separatrix, the linear scaling relation (1) is not satisfied.
Nonetheless, it is striking that the data near the separatrix for
the two subbands lie approximately on the same nonlinear curve.
The simplest interpretation of this apparent scaling behavior is
that these energies probe the dispersion of 2D graphene at a
wavevector $q_n = n/3R$.   This suggests that the ratio problem
and the blue shift problem have the same origin.

Gonzalez et al.\cite{guinea} have shown that the Coulomb
interaction in 2D graphene leads to a singular correction to
the electron self energy.   Consider the Hamiltonian
\begin{equation}
{\cal H} = \hbar v_F \int d^2r \psi^\dagger \vec\sigma\cdot{\vec\nabla\over
i}
\psi+ {e^2\over 2} \int d^2 r d^2 r'
{n({\bf r}) n({\bf r}')\over{|{\bf r}-{\bf r}'|}},
\end{equation}
where $\psi$ is a Dirac spinor with two copies for the
K-K' degeneracy, and $n = \psi^\dagger\psi$.
The Coulomb interaction is characterized by a
dimensionless interaction strength $g = e^2/\hbar v_F$.
In lowest order perturbation theory the electronic dispersion
is
\begin{equation}
E(q) = \hbar v_F q [1 + (g/4)\log(\Lambda/q)],
\end{equation}
where $\Lambda$ is an ultraviolet cut off of order the inverse lattice
constant.   The nonlinear behavior as $q \rightarrow 0$ is a consequence of
the long range singularity of the 2D Coulomb
interaction $V(q) = 2\pi e^2/q$.  It is thus important to account
for screening.  The semi-metallic Dirac spectrum of
graphene leads to a static polarizatibility $\Pi(q) =
(1/4)q/v_F$.  The linear dependence on $q$ exactly cancels the
$1/q$ singularity of $V(q)$, leading to a multiplicative
renormalization of the interaction analogous
to screening in a 3D dielectric.  The $q\rightarrow 0$
logarithmic correction to $E(q)$ survives screening although its
coefficient is renormalized.  In a static
screening approximation the renormalized interaction is $g_{\rm
scr} = g/(1 +  g \pi/2)$.

Though it is derived in for small $g$, this result has deeper implications,
since it shows that the weak interaction limit is perturbatively
stable.  (3) is invariant under the renormalization group (RG) transformation
$\Lambda \rightarrow \Lambda e^{-\ell}$, $g \rightarrow g(\ell)$,
$v_F \rightarrow v_F(\ell)$ with
\begin{equation}
dg/d\ell = -g^2/4; \quad\quad\quad dv_F/d\ell = v_F g/4.
\end{equation}
We may interpret (3) in terms of a scale dependent
renormalization of $v_F$ and $g$.  The interaction vertex $e^2 =
\hbar v_F g$ is not renormalized, so that the scaling is
characterized by a single parameter $g$. Eq. (4) shows that
$g$ is marginally irrelevant: at long wavelengths $g$ becomes
smaller and perturbation theory becomes better. This implies that
even for strong interactions the system flows to the perturbative
limit at long wavelengths where (3) and (4) are valid. Therefore,
the dispersion for small $q$ is given {\it exactly} by (4)
with renormalized parameters $v_F$ and $g$, which depend on the
cutoff scale $\Lambda$.  We thus have a situation
similar to Fermi liquid theory, where low energy
quasiparticles behave like non interacting particles, albeit with
renormalized parameters. Here, however, the marginal irrelevance
of $g$ leads to logarithmic corrections which do not disappear at
low energies.   As emphasized by Gonzalez et al.\cite{guinea},
this singular behavior is a signature of a marginal Fermi liquid.

Though (3) is exact for
$q\rightarrow 0$, it remains to determine the values of the renormalized
parameters $v_F$ and $g$ when the bare interactions are strong.
Using the Fermi velocity of bulk graphite (where 3D
screening eliminates the logarithmic singularity),
$\hbar v^0_F = 5.3$ eV\AA \ we estimate a bare interaction
strength of $g_0 = e^2/\hbar v_F^0 = 2.7$ at a cut off scale
$\Lambda_0$ of order the inverse lattice constant.   A crude
estimate of the renormalized parameters may then be obtained by
extrapolating (4) to strong coupling using $g = (g_0^{-1} +
(1/4)\log \Lambda_0/\Lambda)^{-1}$.  For $\Lambda \sim .5 {\rm
nm}^{-1}$ this gives $g \sim 1.1$ and $\hbar v_F \sim 12.9$
eV\AA. A more accurate theory requires knowledge of the form of
the RG flow equations (4) for strong coupling and requires an
approximation. Gonzalez et al.\cite{guinea} have developed a GW approximation,
which incorporates a dynamically screened Coulomb interaction.  A
simpler theory can be developed within a statically screened
approximation.  We find that the results agree within 5\% with
the dynamically screened theory\cite{disagree}.  For static
screening
 the renormalized dispersion has the same form as (3)
with $g$ replaced by $g_{\rm scr} = g/(1 + g \pi/2)$.  The RG
flow equations are similarly modified with a factor of
$(1+g\pi/2)^{-1}$ on the right hand side of (4).  This leads to a refined estimate of the
parameters at $\Lambda \sim .5 {\rm nm}^{-1}$:  $g = 2.0$ ; $\hbar
v_F = 7.2$ eV\AA.  The screened interaction is $g_{\rm scr} = 0.48$.

The nonlinear scaling form of the separatrix in Fig. 1 is
consistent with Eq. (3).  Choosing the  scale $\Lambda = 0.5 {\rm
nm}^{-1}$, the data is well fit with the parameters $v_F =
7.8$ eV\AA\ and $g = 0.74$.  These parameters are in acceptable
agreement with the statically screened theory described above,
given the theory's simplicity.  The 2D interactions
in graphene appear to explain the nonlinear scaling of the data
in Fig. 1 and thus resolve both the ratio problem and the blue
shift problem.

Nevertheless the agreement between the data and the interacting
theory of 2D graphene is surprising because the
latter does not account for excitonic effects, which are known to
be large\cite{ando,spaturu,tersoff}.  To describe excitons it is essential
to account for the 1D
interactions on scales {\it larger} than $R$.
In addition to exciton binding, these interactions lead to an
increase of the single particle energy gap.  To address this
issue we have numerically calculated both the single particle and
particle-hole gaps.  We find that the two 1D
interaction effects largely cancel one another, so that the
$R$ dependence of the particle-hole gap is ultimately well
described by the 2D theory.   We will begin by
discussing our numerical calculation.  We then show how these
conclusions can be understood within a simple 1D
model.

We have computed the the single particle and particle hole energy
gaps for nanotubes in a statically screened Hartree-Fock
approximation.  Our calculation is similar to that previously
reported by Ando\cite{ando}, though here we focus on the $R$
dependence of the energy gaps.  We use
a $\pi$ electron tight binding model, which includes all 1D subbands.
To avoid the complications
associated with the $\nu\sin3\theta/R^2$ corrections we
study semiconducting tubes by calculating excitations of
armchair tubes with an appropriate phase shifted boundary
condition impose an energy gap. The single
particle band gaps are computed by evaluating the exchange self
energy using a statically screened Coulomb interaction.  The
particle-hole gap is determined by numerically diagonalizing the
Schrodinger equation for the particle and the hole in the
renormalized bands bound by the screened interaction.  This is
equivalent to solving the Bethe-Salpeter equation in the static
screening approximation.

\begin{figure}
 \centerline{ \epsfig{figure=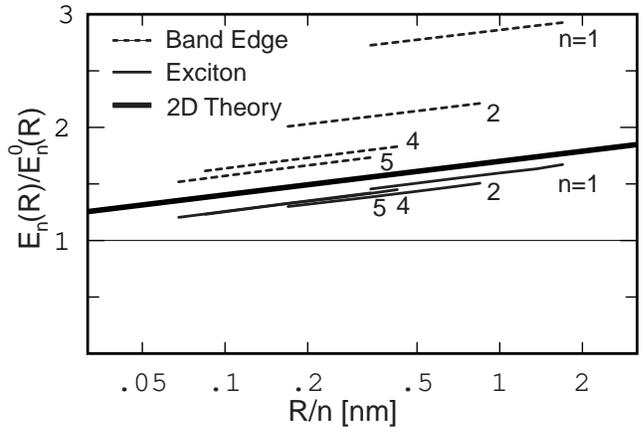,width=3.5in} }
 \caption{Single particle gaps (dashed line) and particle hole gaps (solid line)
  for the first four subbands of semiconducting $[p,p]$ nanotubes
 with phase shifted boundary conditions calculated for $5<p<25$.   The thick line
 is the prediction of the 2D theory Eq. 4. }
 \end{figure}

In Fig. 2 we plot the single particle and particle hole gaps as a
function of radius and subband index.  To emphasize
the corrections to linear scaling behavior we provide a log-linear
plot of $E_n(R)/E_n^0(R)$ as a function of $R/n$, where $E_n^0$
is given by (1), and is proportional to $n/R$. The prediction
based on the statically screened 2D theory of
graphene given in (4) is shown for comparison
\cite{uncertainty}.  The single particle gaps are strongly enhanced relative
to their non interacting values, while the particle-hole gaps are
only moderately enhanced.  Thus, most of the enhancement of the
single particle band gap is cancelled by the electron hole interaction that
binds the exciton. Moreover, since the slopes of all of the curves
is the same in Fig. 2, both the single particle and the particle
hole gaps exhibit the same logarithmic increase with radius.  The
excitonic binding energy, which is the difference between the
two, does not have the logarithmic increase, and scales
inversely with $R$. The particle-hole gaps for the different
subbands lie nearly on a {\it single}  straight
line,  close to the prediction of the 2D interacting
theory. This is consistent with the scaling behavior in the
experimental data in Fig. 1. In contrast, the single particle gaps
are well above the predictions of the 2D theory, and
do not obey scaling with subband index.

The essential features in Fig. (2) can be understood within a
simpler model for the 1D interactions on scales
larger than the tube radius.  For example, consider a
semiconducting nanotube with a bare energy gap $2\Delta$ with an
{\it infinite range} interaction $V(x) = V_0$. This is the
constant interaction model, familiar from the theory of the
Coulomb blockade.  In this model the interaction energy is  $V_0
N^2/2$, where $N$ is the total number of electrons.  The single
particle energy gap is then simply $2\Delta + V_0$.  The
particle-hole energy gap, which determines the energy of optical
transitions is $2\Delta$.  Since the exciton is electrically
neutral, its energy is
unaffected by the infinite range interaction. For this model the
exciton binding energy {\it exactly} cancels the enhancement of
the single particle gap.

Though the 1D Coulomb interaction $V_0(q) = 2 e^2 \log qR$ is not
truly infinite range, the infinite range limit is an appropriate
starting point for describing the 1D effects. In the
static screening approximation, $V_{\rm scr}(q) = V_0(q)/(1
+ V_0(q) \Pi(q))$.  Since the 1D polarizatibility $\Pi(q) \sim
q^2R^2/v_F$ for small $q$, the $q \rightarrow 0$ part of the
interaction is unscreened and screening suppresses only the
shorter wavelength components of the potential. This leaves a
screened interaction which is more strongly peaked at low momenta
$qR<<1$, i.e. closer to the infinite range limit. Note that this
is consequence of the one dimensionality of the nanotube and has
no analog in a 3D semiconductor, where the long
range interaction is uniformly reduced by the dielectric constant.
These considerations help to explain the behavior in Fig. 2. The
net effect of the long range 1D interactions on the
excited states is relatively small in spite of the fact that the
renormalization of the single particle energy gaps and the
binding energy of the electron hole pair are separately quite
strong.

The scaling of the exciton binding energy, $E_B$ with $R$
 may also be considered in a simple 1D model.
Begin with the Hamiltonian (2) defined on a cylinder of radius $R$, and
integrate out the high energy degrees of freedom down to a cutoff
scale $\Lambda \sim 1/R$.  The renormalized Hamiltonian then has the same
form as (2), and depends only on three parameters, $e^2$, $v_F$ and
$R$.  It follows that the eigenvalues of ${\cal H}$
have the scaling form, $(e^2/R) f(g)$, where $g = e^2/\hbar v_F$ is
the interaction at scale $1/R$\cite{irrelevant}.
Since $g$ is scale
dependent, the absence of the logarithmic correction of $E_B$ in Fig.
2 implies that the scaling function $f$ is independent
of $g$.   Perebeinos et al. \cite{tersoff} have found an approximate
scaling relation for the exciton binding energy for interactions screened
by a dielectric constant $4<\epsilon<15$.  For nearly armchair
tubes with effective mass $m \sim 1/(v_F R)$ they find $f(g) \sim
g^{\alpha-1}/\epsilon^\alpha$, with $\alpha \sim 1.4$.  This describes
the crossover between the Wannier limit $\epsilon \gg 1$, where $f(g) \sim g/\epsilon^2$
and a strong interaction limit $\epsilon \sim 1$ where
the dependence on $g$ is weak.

Note that this scaling argument does not imply that the
the band gap renormalization and exciton binding  scale like
$n/R$. The apparent scaling behavior for the particle hole gaps
in Fig. 2 is a consequence of the cancellation between the 1D interaction effects.
Because of this near cancellation, the
effects of the two dimensional electronic interactions can be
seen clearly in the experimental data. It is interesting that
theory presented derived from the leading order contributions in
$1/R$ to the excitation energies provides a good description of
the data over the range of experimentally measured tube radii.

We also note that the large single particle gaps shown in Fig. 2
are likely to be important for many nanotube-derived devices, but
have yet to be measured directly in experiments done to date.
They are accessible in principle by measuring the activation
energy for transport in a semiconducting tube, or by measuring
the threshold for photoconductivity following optical excitation
into the lowest subbands. Interpretation of the gaps measured in
scanning tunneling spectroscopy are complicated by screening
effects from the substrate, and make it difficult to extract the
single particle gap of individual tubes.

We thank Bruce Weisman for helpful discussions.  This work was
supported by the NSF under MRSEC grant DMR-00-79909 and the DOE under
grant DE-FG02-ER-0145118.


\begin{thebibliography}{10}
\bibitem{bachillo} S.M. Bachilo, et al., Science {\bf 298}, 2361
(2002).

\bibitem{lefebvre}  J. Lefebvre, Y. Homma and P. Finnie, Phys. Rev.
Lett. {\bf 90} 217401 (2003).

\bibitem{hagen} A. Hagen and T. Hertel, Nano Lett. {\bf 3} 383 (2003).

\bibitem{ando} T. Ando, J. Phys. Soc. Japan {\bf 66} 1066 (1996).

\bibitem{kanemele} C.L. Kane and E.J. Mele, Phys. Rev. Lett. {\bf 90}
207401 (2003).


\bibitem{spaturu} C.D. Spataru, S. Ismail-beigi, L. Benedict and S.G. Louie,
Phys. Rev. Lett. {\bf 92}, 077402 (2004).

\bibitem{tersoff} V. Perebeinos, J. Tersoff and P. Avouris,
cond-mat/0402091 (2004).

\bibitem{guinea}   J. Gonzalez, F. Guinea and M.A.H. Vozmediano, Phys. Rev. B. {\bf 59},
2474 (1999).

\bibitem{kmcurvature} C.L. Kane and E.J. Mele, Phys. Rev. Lett. {\bf
78}, 1932 (1997).

\bibitem{reich} S. Reich and C. Thomsen, Phys. Rev. B {\bf 62} 4273
(2000).

\bibitem{disagree}  We confirmed the form of the RG flow
equations derived in \onlinecite{guinea} in the GW approximation.
However, we find the equations in \onlinecite{guinea} contain
incorrect multiplicative factors.

\bibitem{uncertainty}  The 2D prediction is only known acurrately up
to a additive factor, since the precise value of the cutoff $\Lambda$
is not specified.  However, the slope is determined by the
interaction strength.

\bibitem{irrelevant}  Irrelevant operators such as
$\psi^\dagger \nabla^2 \psi$ and $(\psi^\dagger\psi)^2$ will also be present.  The latter
leads to a $1/R^2$ correction to the transition energies.  These
effects are small in Fig. 2.


\end{thebibliography}
\end{document}